\documentclass{article}

\usepackage[a4paper, total={6in, 8in}]{geometry}

\usepackage{amsfonts,amsmath}
\usepackage{natbib}
\usepackage{xcolor}

\usepackage{graphicx}
\usepackage{dcolumn}
\usepackage{bm}
\usepackage[]{hyperref}
\hypersetup{colorlinks=true, allcolors=blue!50!black}
\usepackage[capitalise]{cleveref}
\creflabelformat{equation}{#2#1#3}

\usepackage[makeroom]{cancel}

\newcommand{\nameDisorder}{{temporal path entropy}}
\newcommand{\NameDisorder}{{Temporal path entropy}}

\title{Higher-Order Patterns Reveal Causal Timescales \\ of Complex Systems}

\usepackage{authblk}
\author[1]{Luka V. Petrovi{\'c}}
\author[2]{Anatol Wegner}
\author[1,2]{Ingo Scholtes}
\affil[1]{Data Analytics Group, University of Zurich, Z\"{u}rich, Switzerland}
\affil[2]{Center for Artificial Intelligence and Data Science (CAIDAS), Julius-Maximilians-Universit\"at W\"urzburg, W\"urzburg, Germany}

\date{\today}

\begin{document}

\maketitle

\begin{abstract}
The analysis of temporal networks heavily depends on the analysis of time-respecting paths.
However, before being able to model and analyze the time-respecting paths, we have to infer the timescales at which the temporal edges influence each other. 
In this work we introduce \nameDisorder{}, an information theoretic measure of temporal networks, with the aim to detect the timescales at which the causal influences occur in temporal networks.
The measure can be used on temporal networks as a whole, or separately for each node.
We find that the \nameDisorder{} has a non-trivial dependency on the causal timescales of synthetic and empirical temporal networks.
Furthermore, we notice in both synthetic and empirical data that the \nameDisorder{} tends to decrease at timescales that correspond to the causal interactions.
Our results imply that timescales relevant for the dynamics of complex systems can be detected in the temporal networks themselves, by measuring \nameDisorder{}.
This is crucial for the analysis of temporal networks where inherent timescales are unavailable and hard to measure.
\end{abstract}

\section{Introduction}

The research of dynamic complex systems has in recent years advanced beyond static graph representations \citep{lambiotte2019networks,battiston2020networks}.
The focus has shifted to various generalizations of diadic interactions in graphs: multiple types of interactions in multilayer network~\citep{kivela2014multilayer}, multibody interactions in the form of simplicial complexes and hypergraphs \citep{petri2018simplicial} and models that incorporate concepts of memory \citep{scholtes2014causality,lambiotte2015effect,williams2022shape}.
Such generalized relationships allow us to better model complex systems, because they can represent richer data.

Temporal networks are one kind of such rich data; they record not only who interacted with whom, but also when each interaction happened.
They bring us closer to understanding the dynamics of complex systems, but require us to perform analysis beyond the static networks approach~\citep{holme2012temporal,holme2015modern}. 
The time information can yield valuable insights on its own~\citep{goh2008burstiness}, and, although initially the temporal and topological aspects of temporal networks were mostly studied independently, even richer insights are hidden in the coupling of the temporal and topological patterns~\citep{ceria2022topological}.
Such coupling can affect the statistics of time-respecting paths~\citep{holme2012temporal} in temporal networks, which impacts e.g., analysis of accessibility~\citep{lentz2013unfolding}, reachability~\citep{badie2020efficient}, spreading~\citep{masuda2013temporal,scholtes2014causality,lambiotte2015effect,badie2022directed}, clustering~\citep{rosvall2014memory}, centralities~\citep{scholtes2016higher}, and visualization~\citep{perri2020hotvis}.
Although there are many possible ways in which temporal and topological patterns can couple in complex systems, one of the most basic cases is when an incoming temporal edge to a node causes a change of frequencies of the edges emanating from the node within a given time-window.
For instance, in a communication network we expect an incoming message to induce a outgoing message on the same topic, e.g. in the form of a reply, within a certain time window reflecting the minimal reaction time and memory of the recipient.
However, information on the timescales relevant for the temporal network dynamics is rarely available in real world settings.

In this work, we define an information theoretic measure to detect timescales at which interactions in a complex system cause each other.
We demonstrate its effectiveness in both synthetic and real world data.

\section{Background}

\newcommand{\real}{\mathbb{R}}

\newcommand{\deltatmin}{{\tau_{\min}}}
\newcommand{\deltatmax}{{\tau_{\max}}}
\newcommand{\timewindow}{{\tau}}

\newcommand{\deltatminGT}{{\bar{\tau}_{\min}}}
\newcommand{\deltatmaxGT}{{\bar{\tau}_{\max}}}
\newcommand{\timewindowGT}{ {\bar{\tau}}}

\newcommand{\tempnet}{\Gamma}
\newcommand{\nodes}{{V}}
\newcommand{\edges}{{E}}
\newcommand{\tedges}{{\mathcal{E}}}

\newcommand{\paths}{{ \mathcal{P} }}
\newcommand{\disorder}{{\mathcal{H} }}

\newcommand{\counts}[1]{{n_{#1}}}

\newcommand{\TotalTime}{{T_{\text{total}}}}
\newcommand{\causalpaths}{{ \mathcal{P}_{\text{causal}} }}

Let $\tempnet= (\nodes, \tedges)$ be a temporal network consisting of a set of nodes $\nodes$ and a set of time-stamped edges $\tedges \subseteq \nodes \times \nodes \times \real$.
We denote the set of unique edges with $\edges \subset \nodes \times \nodes$.
A temporal edge $(v,w,t)\in \tedges$ represents a direct link from node $v$ to node $w$ at time $t$.
For simplicity, we assume that the temporal edges are instantaneous, however the method and algorithms can be modified in a straightforward fashion to the case where edges have finite duration.\textbf{}
Formally, we call a sequence of time-stamped edges  $(v_1,w_1,t_1)$, \ldots, $(v_{k},w_{k},t_{k})$ a time-respecting path iff for all $ i \in \{2, \ldots, k\}$ they satisfy the following conditions~\citep{pan2011path,holme2012temporal,casteigts2021finding}:
\begin{align}
    w_{i-1} &= v_{i} \label{tsd:eq:time-respected-path-1} \\
    \deltatmin &< t_{i} - t_{i-1} < \deltatmax. \label{tsd:eq:time-respected-path-2}
\end{align}
The parameters $\deltatmin$ and $\deltatmax$ naturally introduce a timescale that affects all analyses of temporal networks that are based on time-respecting paths.

The timescale has to be defined differently for processes \emph{on} the temporal network or the processes \emph{of} the temporal network~\citep{holme2012temporal}.
In the former case, the timescale is defined by the process running on the temporal network, e.g. in the case of an epidemic that is spreading over a temporal network of contacts, the timescale is a property of a disease, related to the time interval in which a person is contagious and not related to the timescales at which contacts occur~\footnote{We note that the processes on and of the temporal network may interact~\citep{gross2009adaptive}, and thus blur the distinction. }.
In the latter case, the timescale is part of the process of edge activation, and thus shapes the temporal network itself.
For example, information that is spreading between individuals is also affecting the individuals' choice with whom to share the information: a person would be more likely to share the family-related information with a family member and work-related information with a colleague.
In this letter, we investigate the latter case, more specifically, we investigate whether interactions in a complex system induce one another at a given timescale $\timewindow = [\deltatmin,\deltatmax]$.

In the literature, there exist a variety of definitions of timescales in temporal networks, as well as a variety of methods aimed at detecting them.
The various definitions of timescales are based on the different structural features of temporal networks.
One popular definition of timescales in temporal networks is the approach based on splitting the network into time-slices and aggregating the edges inside the time-interval~\citep{caceres2013temporal,darst2016detection}.
In the same framework, \citet{ghasemian2016detectability} and \citet{ taylor2016enhanced} investigate the limitations of detectability of cluster structures dependent on the timescales of aggregation.
Since this framework is based on aggregating the temporal network into a sequence of static time-aggregated networks, it loses information of the time-respecting paths and is therefore not in line with our aims.
Other lines of research often related to timescale detection are change point detection ~\citep{peixoto2018change}, and analysis of large-scale structures.
\citet{gauvin2014detecting} detects clusters and their temporal activations in a temporal network using tensor decomposition.
Similarly, \citet{peixoto2015inferring} proposed a method to detect the change points of cluster structure in a temporal network.
\citet{peixoto2017modelling} proposed a method to simultaneously detect the clusters and timescales in temporal network, however, they model the temporal network as a single sequence of tokens (similar to~\citep{peixoto2018change}) that represent temporal edges, and their timescale inference refers to the number of tokens in the memory of a Markov chain that models such a sequence.
In our view, these works focus on mesoscale structures, and take a coarse grained view of temporal networks, while in this work, we propose a complementary approach by focusing on local correlations between temporal edges incident on a node and subsequent temporal edges emanating from it.
Among the works that took a fine-grained view, \citet{williams2022shape} investigated pairwise correlations between the temporal edges.
Different from the approach that we took, they aggregate the network in time-slices as a preprocessing step, and the timescale is defined as a maximum number of time slices back in time at which correlations are detectable.
\citet{scholtes2016higher} found that correlations between edges on time-respecting paths affect centralities; they modeled the time-respecting paths with higher-order models and found that this approach improves the centrality rankings.
They identified the issue of timescale detection in the context of time-respecting paths, which our work addresses.
Our work also complements the work of \citet{pfitzner2013betweenness} which introduces betweenness preference that can be used to study over- and under-represented time-respected paths in temporal networks, but does not address the problem of detecting the timescales at which these paths occur.
To the best of our knowledge, our work is the first to address the issue of timescale detection for time-respecting paths in temporal networks.

\section{Temporal Path Entropy}

We address the issue of timescale detection by analysing the statistics of time-respecting paths $\paths_\timewindow^k(\tempnet)$ of length $k$ at timescales $\timewindow = [\deltatmin, \deltatmax]$ in a temporal network $\tempnet$.
We define ``\nameDisorder{}'' $\disorder$ for paths $(v_0, v_1, \ldots, v_{k})$ of length $k$ as the entropy of the last node $v_k$ conditional on the sub-path $(v_0, v_1, \ldots, v_{k-1})$:
\begin{align}
    \disorder &= H( v_k | v_0, \ldots, v_{k-1}) \label{tsd:eq:def-disorder}  \\
    &= H(v_0, \ldots, v_k) - H(v_0, \ldots, v_{k-1}),
     \label{tsd:eq:disorder-rewritten}
\end{align}
where $H(P)=-\sum_i p_i\ln(p_i)$ is the Shannon entropy.
The identity in \cref{tsd:eq:disorder-rewritten} can be obtained by applying the chain rule (see Appendix for derivation).
By definition, \nameDisorder{} $\disorder$ measures uncertainty in the last step of time-respecting paths given the $k-1$ previous steps.
A lower value of the entropy indicates a high correlation between the memory of time-respecting paths and subsequent steps.
Hence the $\timewindow$ for which the entropy reaches its minimum gives us the timescale for which time-respecting paths become most predictable, i.e. where the correlations between subsequent temporal edges are the most pronounced.
The entropy can also be defined for a single node $v$, by simply fixing $v_{k-1} = v$, allowing for a more fine grained analysis that could be important if nodes differ significantly with respect to the timescales they operate on. 
The intuition behind the \nameDisorder{} is to measure how much the target $v_k$ of an edge emanating form the node $v_{k-1}$ depends on the incoming paths that influenced it in the past.
Testing those dependencies at different timescales would thus point to the timescales at which the dependencies are most pronounced. 
When we compute \nameDisorder{} for the whole temporal network, we use all time-respecting paths of length $k$ in the temporal network, while when we compute it for a node $v$, we select only the paths where $v_{k-1} = v$.

\begin{figure*}[!ht]
    \includegraphics[width = \textwidth]{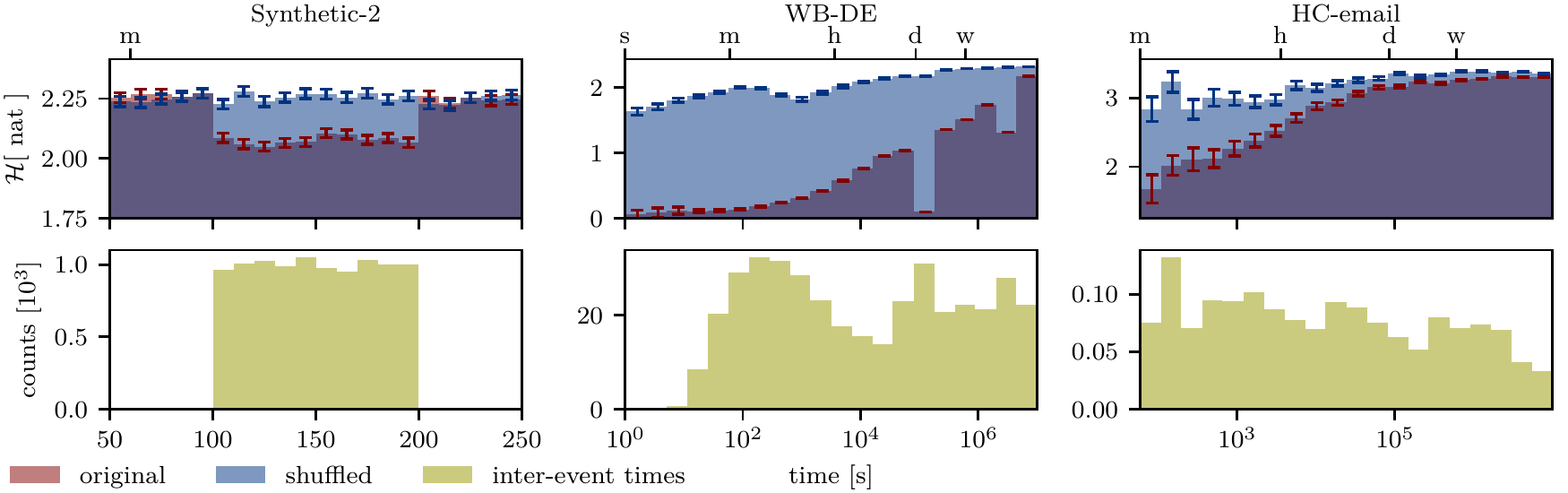}
    \caption{
    Top: \nameDisorder{} as a function of causal temporal scales in datasets Synthetic-2, WB-DE, and HC-email (transparent red) and in the temporal networks with shuffled timestamps (transparent blue).
    The height of a bar represents \nameDisorder{} (error bars represent the estimation error) and the $x$-limits of a bar represent the interval $ \timewindow = [\deltatmin,\deltatmax]$ on which the \nameDisorder{} was measured.
    We indicate on $x$-axis the timescales of one minute (m), hour (h), day (d), week (w), and year (y).
    We observe that the \nameDisorder{} differs more between the original and the shuffled network at causal timescales.
    Bottom: histogram of causal inter-event times.
    }
    \label{tsd:fig:results}
\end{figure*}
In practice, the \nameDisorder{} can be estimated from the counts of time-respecting paths $\paths_\timewindow^k(\tempnet)$ by assuming multinomial distributions with respective probabilities $p( v_0, \ldots, v_{k-1})$ and $p( v_0, \ldots, v_{k})$.
The counts of time-respecting paths can be computed e.g. using the methods from ~\citep{kivela2018mapping,petrovic2021paco}. 
The estimation of the entropy can be challenging especially for small ranges of timescales, since the temporal network can get temporally disconnected, resulting in very few paths of order $k$ being observed.
As a result we require an efficient method for estimating the entropy that performs well even in such under-sampled regimes.
The simplest estimator of a multinomial distribution, called the plug-in estimator, is based on the maximum likelihood estimation, which, however, is known to severely underestimate the entropy in the undersampled regime and has various corrections \citep[e.g.][]{miller1955note, grassberger2003entropy}).
An alternative to the plug-in estimator is to follow a Bayesian approach which results in entropy estimators that strongly depend on the choice of prior.
To counteract this dependency, the NSB estimator~\citep{nemenman2001entropy}  directly infers the entropy from the counts by averaging over different priors for the transition probabilities, rather than inferring the transition probabilities themselves.
Being a Bayesian method, the NSB estimator can also be used to quantify the uncertainty of the estimates. 
Assuming that the estimates of $H(v_0,\ldots, v_k)$ and $H(v_0,\ldots, v_{k-1})$ have independent errors $\sigma_k$ and $\sigma_{k-1}$, we can approximate the total error of the estimate as $\sigma = (\sigma_{k}^2 + \sigma_{k-1}^2 )^{1/2}$.
As the NSB estimator requires the size of the alphabet to be known, it is most suitable for cases where the number of nodes is fixed and improves further if the set of edges that can occur are known a priory as this further restricts the number of potential paths.
In cases when the number of nodes in the system is unknown, the Pitman-Yor Mixture entropy estimator~\citep{archer2014bayesian} could be used instead.

Finally, we address testing whether an interval $\timewindow$ is a causal timescale of a temporal network $\tempnet$.
To do so, we need to assume the null hypothesis that there are no temporal correlations between temporal edges, but the main issue is to obtain a sample of temporal networks under this assumption.
To resolve this issue, we repeatedly randomize the observed temporal network $\tempnet$ by randomly permuting timestamps between its temporal edges~\citep{holme2012temporal}.
These samples of temporal networks would preserve both the edge frequencies and timestamp distribution while destroying the correlations between temporal edges. 
We can use the samples to determine whether \nameDisorder{} of the observed network has an unexpected value under the null assumption.

\section{Experiments}

In the following part, we first show the behavior of \nameDisorder{} in synthetically generated temporal networks with known causal timescales (the description of the generation process can be found in the Appendix); we then present how it behaves in two real-world networks with the information about the ground truth timescales, and two real world networks without the information about the ground truth timescales.

In the top left panel of \cref{tsd:fig:results}, we present the \nameDisorder{} $\disorder$ ($y$-axis) for various timescales ($x$-axis): the left and right $x$ limit of a bar represents $\deltatmin$ and $\deltatmax$, and the height of the bar represents $\disorder$.
The results are shown both for the synthetic network and its shuffled network.
In the bottom left panel, we show the histogram of inter-event times on causal paths.
We observe that the \nameDisorder{} behaves as expected and decreases in accordance with the planted timescale at which the interactions cause one another.
Moreover, this pattern disappears when the timestamps of edges are shuffled, demonstrating that \nameDisorder{} captures the interplay of temporal and topological patterns.

\begin{figure}
    \centering
    \includegraphics[]{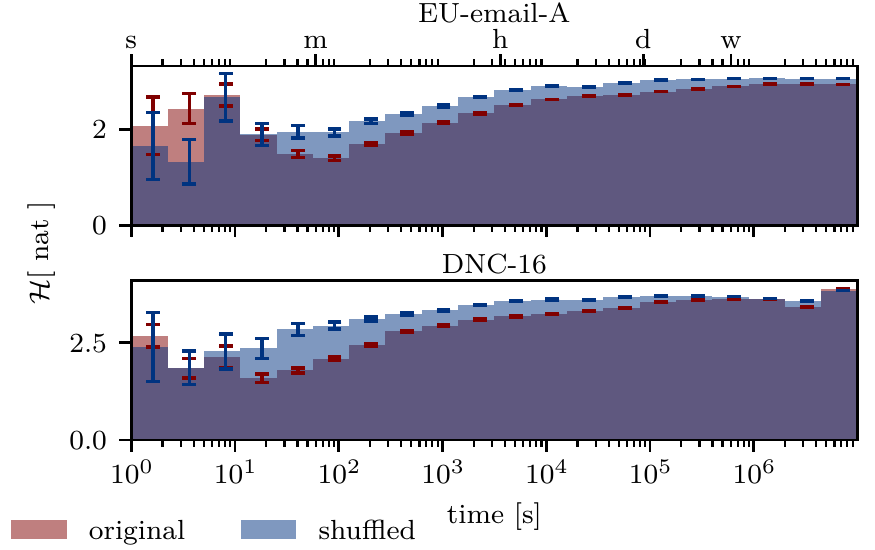}
    \caption{\NameDisorder{} as a function of the timescale $\timewindow$ in EU-email-A and DNC-16 and in timestamp shuffled networks. Timescale $\timewindow$ is represented with the $x$-limits of the bar, and \nameDisorder{} is represented as the height of the bar. Error bars indicate the error of the \nameDisorder{} estimates.}
    \label{tsd:fig:fig_diff_vs_shuffle1_dnc_eu}
\end{figure}

We consider here two empirical temporal networks where we have information about the ground truth causal structure and two empirical temporal networks where we have no information about the ground truth causal structure.
As a first data set, we consider the bipartite temporal network of German Wikibooks co-editing patterns (WB-DE)~\citep{edit_wikibooks,netzschleuder}.
This data contains information about edits on the Wikibooks website: for each edit, we know the editor, the article that was edited, and the time at which the edit occurred.
We preprocess this data to obtain a temporal network of editors: if editor $v$ edited an article prior to editor $w$ who edited the same article at time $t$, we assume that a link $(v,w,t)$ occurred in the temporal network of editors.
We define causal inter-event times based on the articles: we extract the time intervals between successive edits of each article. 
In WB-DE data, we analyze the timescales of the whole temporal network.
As a second data set, we consider public data set of Hillary Clinton's emails (HC-email)~\citep{clintonEmail}, which contains the sender, the receiver, the timestamp, and the subject of each email.
In this data set we analyze the timescales of node representing Hillary Clinton.
While sender, receiver and the timestamp constitute a temporal network, email subjects allow us to obtain causal inter-event times: for each incoming email, we extract the time duration until an email with the same subject was sent.
We use the inter-event times between emails with the same subject and the inter-event times of articles for evaluation; the temporal networks contain only the temporal edges and not any additional information about the ground truth timescales.
We also use two email data sets without a ground truth timescales: EU-email-A \citep{paranjape2017motifs}, which contains email correspondence between researchers of an EU institution from four deparments, and DNC-16 \citep{DNCemail}, which contains emails of the US Democratic National Committee.
Results on other datasets as well as details of all datasets are in the Appendix.
Reproducibility package is available at~\citep{petrovic2023reproducibility}.

Results of the WB-DE and HC-email data are in \cref{tsd:fig:results} (middle and right, respectively).
When we compare the histogram of causal inter-event times  with the \nameDisorder{} at different timescales of the temporal network, we see that increased number of causal interactions increases the difference in \nameDisorder{} between the original and the shuffled network.
The \nameDisorder{} converges for large timescales because the interval sizes increase, the density of causal interactions decreases, and the noise increases.
In \cref{tsd:fig:fig_diff_vs_shuffle1_dnc_eu}, although we do not have the ground truth, we see that the largest difference between the original and the shuffled datasets are at timescales between a minute a day, which is what we would expect from email correspondence.

We identify four limitations of our approach.
First, our base assumption is that the interactions, represented by edges, cause one another, and our measure can not separate that case that from the case when edges are generated by some common factor.
Second, being based on directed paths the current method is restricted in the types of causal interactions it considers namely interactions where a incoming link into  a vertex effects the subsequent links emanating from the vertex.
The method could potentially be generalized to other types of interactions by considering other patterns to alleviate this shortcoming.
Third, our method cannot detect timescales at which the incoming edges to a node change the overall activity of the node without changing the relative frequencies of the outgoing edges.
Detecting timescales of such causal influences is thus an open problem.
Fourth, real data can contain time-varying timescales, e.g. during day or night, which would probably require an application of time warping techniques.

\section{Conclusion}

To summarize, the analysis of temporal networks heavily depends on the analysis of time-respecting paths~\citep{holme2012temporal,holme2015modern,pan2011path,masuda2013temporal,scholtes2016higher,kivela2018mapping}.
However, in order to model and analyze the time-respecting paths, we first need to identify the correct timescale. 
In this work we address this problem by introducing an information theoretic measure, the \nameDisorder{}, that is able to can identify timescales at which the influences are highly correlated.
Using real world data we demonstrated that the measure can be applied to temporal networks as a whole as well as to a single node. 
We showed that the \nameDisorder{} can capture the causal timescales in both synthetic and empirical temporal networks. 
We  further support our findings by observing that the differences in the \nameDisorder{} between the original and shuffled networks coincide with increases in the number of causal paths.
The \nameDisorder{} allows system-relevant timescales to be inferred from the temporal networks themselves which is crucial for the analysis of temporal networks where inherent timescales are unavailable and hard to measure.

\subsection*{Acknowledgments}
    The authors would like to thank Christopher Bl\"ocker, Chester Tan, and Franziska Heeg for valuable comments on the manuscript.
    LP and IS acknowledge support by the Swiss National Science Foundation, grant 176938.
    
\bibliography{refs}

\section{Datasets}

In this work we considered synthetic and empirical temporal networks.

To generate synthetic temporal networks Synthetic-1, Synthetic-2 and Synthetic-3 with a ground truth timescale $ {\timewindowGT} = \left[ \deltatminGT, \deltatmaxGT \right]$, we start from a static Erd\H{o}s-R\'enyi random graph with $50$ nodes and $500$ directed edges.
We sample a random subset $\causalpaths$ of $n_{\text{u.p.}} = 500$ unique paths of length $k = 2$ in the static network which correspond to causal influences in the system.
We sample with repetition  $n_{\text{p}} = 5 000$ paths from $\causalpaths$ to generate dataset Synthetic-1, $n_{\text{p}} = 10 000$ paths to generate dataset Synthetic-2 and $n_{\text{p}} = 20 000$ paths to generate dataset Synthetic-3.
To add each path $(v_0, v_1, v_2)$ to the temporal network, we sample a random starting time $t$ uniformly from $\left[ 0, \TotalTime - \deltatmaxGT \right]$ and create a temporal edge $(v_0, v_1, t)$; we then sample temporal distance $\delta$ between edges on the path (inter-event time) uniformly from $\timewindowGT$ and create the temporal edge $(v_1,v_2, t+\delta)$.
We choose $\timewindowGT$ with $\deltatminGT = 100$ and $\deltatmaxGT = 200$.
To add some noise to the system, we uniformly sample $20 000$ edges from the static graph, and sample their timestamps uniformly from $\left[0,\TotalTime\right]$.
The temporal network Synthetic-4 contains two time-scales relevant for the dynamics.
To do so, we generated two different temporal networks based on two random graphs of $50$ nodes (with the same node names) and $500$ edges and based on the different timescales $\timewindow^1 = [50,100]$ and $\timewindow^2 = [150,200]$.
We used the same procedure as above with parameters $n_\text{u.p.} = 500$; $n_\text{p} = 5000$; $\TotalTime = 10^5$; $n_\text{r.e.} = 10000$.
We merged the two temporal networks into one; the details of the resulting network are in \cref{tsd:tab:datasets}.
The dataset Synthetic-5 contains paths of length three.
Again, there are 50 nodes and 500 edges in the static Erd\H{o}s R\'enyi graph.
We sample $n_{\text{u.p.}} = 20$ unique paths, we sample $n_\text{p} = 20000$ of them, and spread them across $\TotalTime=10^5$ using the same procedure and timescale $\timewindow = [100,200]$.
We add $n_\text{r.e}=10 000$ random edges to the network as noise.

\begin{table}[h]
    \centering
    \begin{tabular}{lrrrr}
    \hline
         dataset &  $|\nodes|$ &  $|\edges|$ &  $|\tedges|$ &  $ \TotalTime \left[s\right]$ \\
    \hline
        Ants-1-1 &          89 &         947 &         1911 &                      1.44e+03 \\
        Ants-1-2 &          72 &         862 &         1820 &                      1.75e+03 \\
        Ants-2-1 &          71 &         636 &          975 &                      1.44e+03 \\
        Ants-2-2 &          69 &         769 &         1917 &                       1.8e+03 \\
        Ants-3-1 &          11 &          37 &           78 &                      1.13e+03 \\
        Ants-3-2 &           6 &          21 &          104 &                      1.42e+03 \\
          DNC-16 &        1891 &        5598 &        39264 &                      8.49e+07 \\
      EU-email-1 &         309 &        3031 &        61046 &                      6.94e+07 \\
      EU-email-2 &         162 &        1772 &        46772 &                      6.94e+07 \\
      EU-email-3 &          89 &        1506 &        12216 &                      6.93e+07 \\
      EU-email-4 &         142 &        1375 &        48141 &                      6.94e+07 \\
      EU-email-A &         986 &       24929 &       332334 &                      6.95e+07 \\
         Gallery &       10972 &       89034 &       831824 &                      6.95e+06 \\
        HC-email &         326 &         385 &         8313 &                      1.19e+08 \\
        Hospital &          75 &        2278 &        64848 &                      3.48e+05 \\
       Hypertext &         113 &        4392 &        41636 &                      2.12e+05 \\
             OSS &        5789 &        6888 &        12583 &                      3.54e+08 \\
         Primary &         242 &       16634 &       251546 &                      1.17e+05 \\
       School-13 &         327 &       11636 &       377016 &                      3.64e+05 \\
     Synthetic-1 &          50 &         500 &        30000 &                         1e+05 \\
     Synthetic-2 &          50 &         500 &        40000 &                         1e+05 \\
     Synthetic-3 &          50 &         500 &        60000 &                         1e+05 \\
     Synthetic-4 &          50 &         898 &        40000 &                         1e+05 \\
     Synthetic-5 &          50 &         500 &        50000 &                         1e+05 \\
           WB-AR &        1124 &        3334 &        27166 &                      3.89e+08 \\
           WB-DE &       10999 &       54700 &       464089 &                      4.87e+08 \\
           WB-FR &        9735 &       53606 &       362094 &                      4.88e+08 \\
         Work-13 &          92 &        1510 &        19654 &                      9.88e+05 \\
    \hline
    \end{tabular}
    \caption{The sizes of the sets of nodes $V$, unique edges $E$, and temporal edges $\tedges$ of temporal networks that we analyzed in the experiments. Datasets synth-2, HC email and WB DE are in the main paper. The other datasets are shown in the Appendix. }
    \label{tsd:tab:datasets}
\end{table}

We also use empirical dataset where can get access to the ground truth causal path structure.
We consider the bipartite temporal network of Wikibooks co-edits in Arabic (WB-AR), French (WB-FR) and  German (WB-DE)~\citep{edit_wikibooks,netzschleuder}.
This data contains information about edits on the Wikibooks website: for each edit, we know the editor, the article that was edited, and the time at which the edit occurred.
We preprocess this data to obtain a temporal network of editors: if editor $v$ edited an article prior to editor $w$ who edited the same article at time $t$, we assume that a link $(v,w,t)$ occurred in the temporal network of editors.
We define causal inter-event times based on the articles: we extract the time intervals between successive edits of each article. 
In these data, we analyze the timescales of the whole temporal network.
Another dataset where we can get access to the ground truth causal structure is the public data set of Hillary Clinton's emails (HC-email)~\citep{clintonEmail}, which contains the sender, the receiver, the timestamp, and the subject of each email.
In this data set we analyze the timescales of node representing Hillary Clinton.
While sender, receiver and the timestamp form a temporal network, email subjects allow us to obtain causal inter-event times: for each incoming email, we extract the time duration until an email with the same subject was sent.
We use the inter-event times between emails with the same subject and the inter-event times of articles for evaluation; the temporal networks contain only the temporal edges and not any additional information about the ground truth timescales.
The details of each data-set are in \cref{tsd:tab:datasets}.

Finally, we also use empirical temporal networks where we do not know the ground truth causal path structure.
Datasets Ants-1-1, Ants-1-2, Ants-2-1, Ants-2-2, Ants-3-1, and Ants-3-3 \citep{blonder2011time} contain antenna contacts in ant colonies. 
Dataset DNC-16 \citep{DNCemail} contains emails of the US Democratic National Committee leaked in 2016.
Datasets EU-email-1, EU-email-2, EU-email-3, EU-email-4, and EU-email-A \citep{paranjape2017motifs} contain email correspondence between researchers of an EU institution from first, second, third, fourth and all departments, respectively.
Datasets Gallery~\citep{isella2011s}, Hospital~\citep{vanhems2013estimating}, Hypertext~\citep{isella2011s}, Primary~\citep{gemmetto2014mitigation,stehle2011high}, Work-13~\citep{genois2015data} and School-13~\citep{mastrandrea2015contact} contain human face-to-face interactions in different settings measured by the SocioPatterns collaborations.
Dataset OSS \citep{zanetti2013categorizing} contains ASSIGN relationships between members of the Open Source Software community Apache.

\section{Results: Synthetic Data}

We present results for datasets Synthetic-1, Synthetic-3, Synthetic-4, Synthetic-5.

\begin{figure}[!ht]
    \centering
    \includegraphics{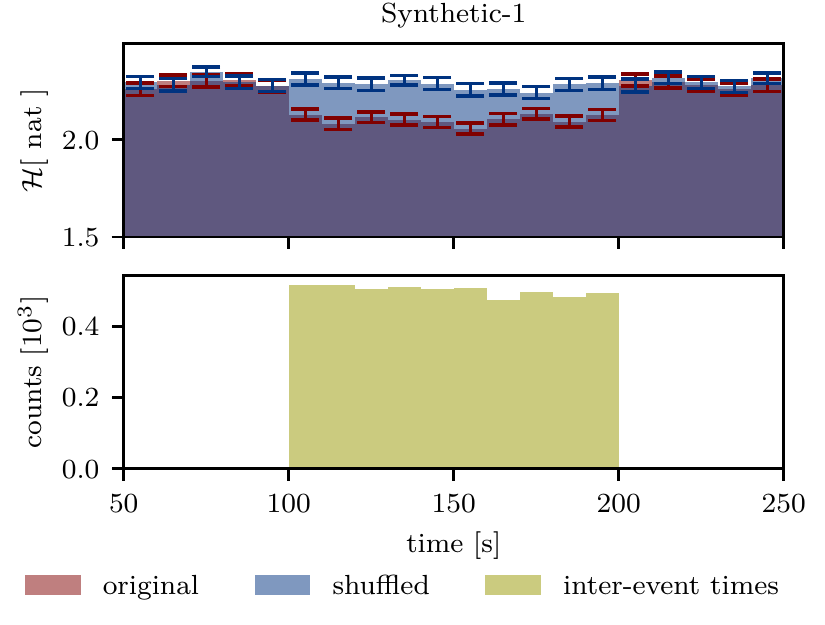}
    \caption{Top: \nameDisorder{} as a function of the timescale $\timewindow$ in temporal network Synthetic-1 and in Synthetic-1 with shuffled timestamps. Timescale $\timewindow$ is represented with the $x$-limits of the bar, and \nameDisorder{} is represented as the height of the bar. Error bars indicate the error of the \nameDisorder{} estimates.
    Bottom: histogram of inter-event times of synthetic causal interactions.
    }
    \label{tsd:fig:fig_diff_vs_shuffle1_synth1}
\end{figure}

\begin{figure}[!ht]
    \centering
    \includegraphics{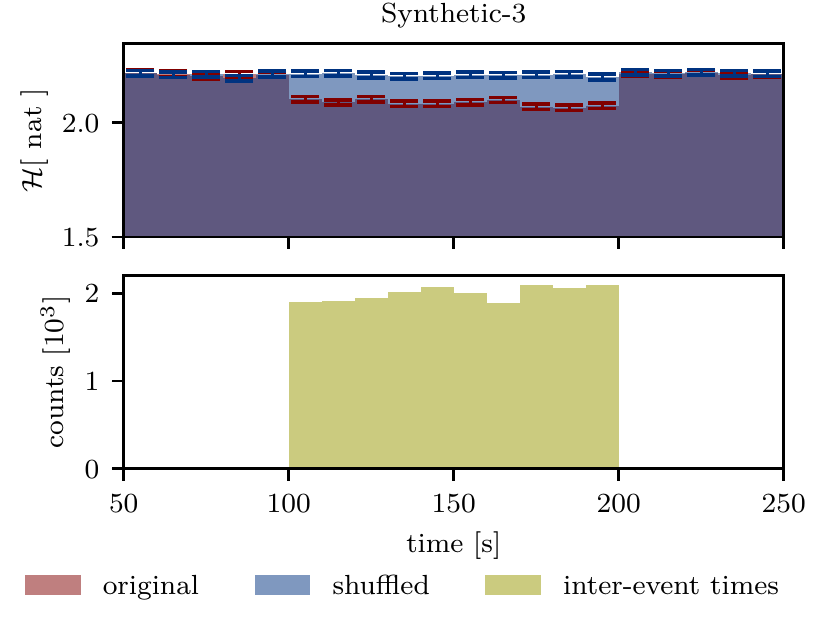}
    \caption{Equivalent of \cref{tsd:fig:fig_diff_vs_shuffle1_synth1}, for Synthetic-3.
    }
    \label{tsd:fig:fig_diff_vs_shuffle1_synth3}
\end{figure}

\begin{figure}[!ht]
    \centering
    \includegraphics{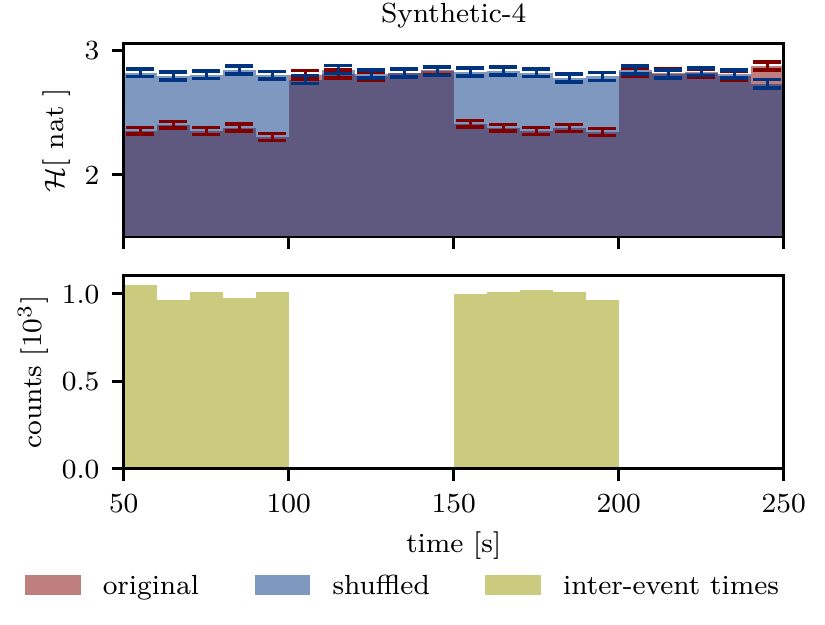}
    \caption{Equivalent of \cref{tsd:fig:fig_diff_vs_shuffle1_synth1}, for Synthetic-4.
    }
    \label{tsd:fig:fig_diff_vs_shuffle1_synth4}
\end{figure}

\begin{figure}[!ht]
    \centering
    \includegraphics{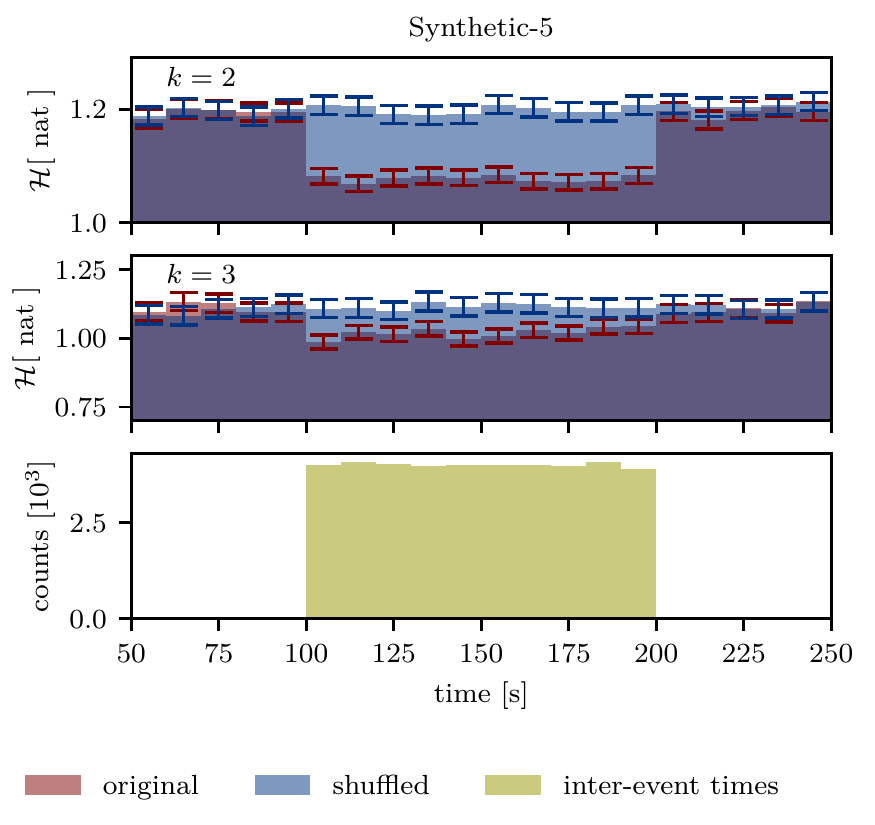}
    \caption{\NameDisorder{} as a function of the timescale $\timewindow$ in temporal network Synthetic-5 and in Synthetic-5 with shuffled timestamps for orders $k=2$ (top) and $k=3$ (middle). Timescale $\timewindow$ is represented with the $x$-limits of the bar, and \nameDisorder{} is represented as the height of the bar. Error bars indicate the error of the \nameDisorder{} estimates.
    Bottom: histogram of inter-event times of synthetic causal interactions.
    }
    \label{tsd:fig:fig_diff_vs_shuffle1_synth5}
\end{figure}

\section{Results: Empirical Data with Ground Truth}
\begin{figure}[t]
    \centering
    \includegraphics[]{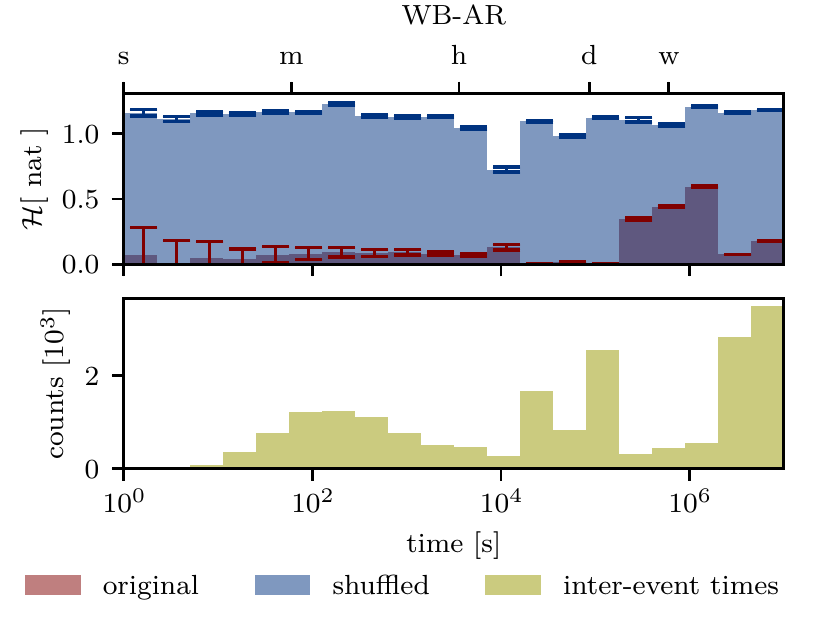}
    \caption{ 
    Top: \nameDisorder{} as a function of the timescale $\timewindow$ in WB-AR temporal network and of WB-AR temporal network with shuffled timestamps. Timescale $\timewindow$ is represented with the $x$-limits of the bar, and \nameDisorder{} is represented as the height of the bar. Error bars indicate the error of the \nameDisorder{} estimates.
    Bottom: histogram of inter-event times for all articles of edits of the same article.
    }
    \label{tsd:fig:appendix-wiki-ar}
\end{figure}

\begin{figure}[t]
    \centering
    \includegraphics[]{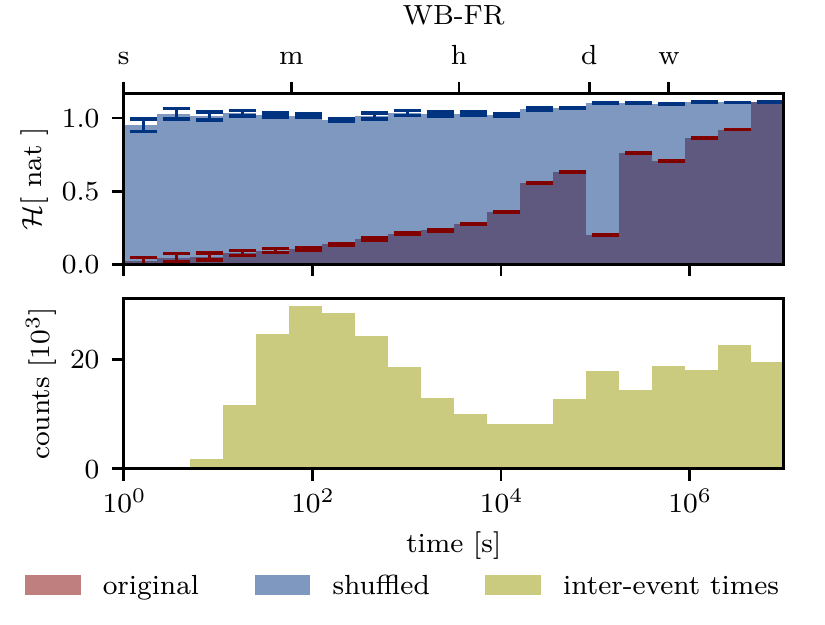}
    \caption{ 
    Equivalent of \cref{tsd:fig:appendix-wiki-ar} for WB-FR.
    }
    \label{tsd:fig:appendix-wiki-fr}
\end{figure}

In this section we show results on other Wikibooks datasets that we used to test the method.
In \cref{tsd:fig:appendix-wiki-ar}, we test \nameDisorder{} on the WB-AR dataset, and in \cref{tsd:fig:appendix-wiki-fr}, we test it on the WB-FR dataset.
Similar to the WB-DE in the main paper, the bottom panel shows the yellow histogram of inter-event times of edits per article for all articles.

\section{Empirical data without the ground truth}

In this section, we show multiple datasets in which we do not have access to the ground truth temporal scales.
Although the lack of ground truth in these datasets makes objective evaluation of the method difficult, the results across datasets are consistent and in accordance with what one would expect: e.g. in the email datasets, \nameDisorder{} is different between the original and the shuffled network for timescales between one minute and a few days, which corresponds to what we would expect to be the interval in which emails are responded to.

\begin{figure*}[!ht]
    \centering
    \includegraphics[width=0.49\textwidth]{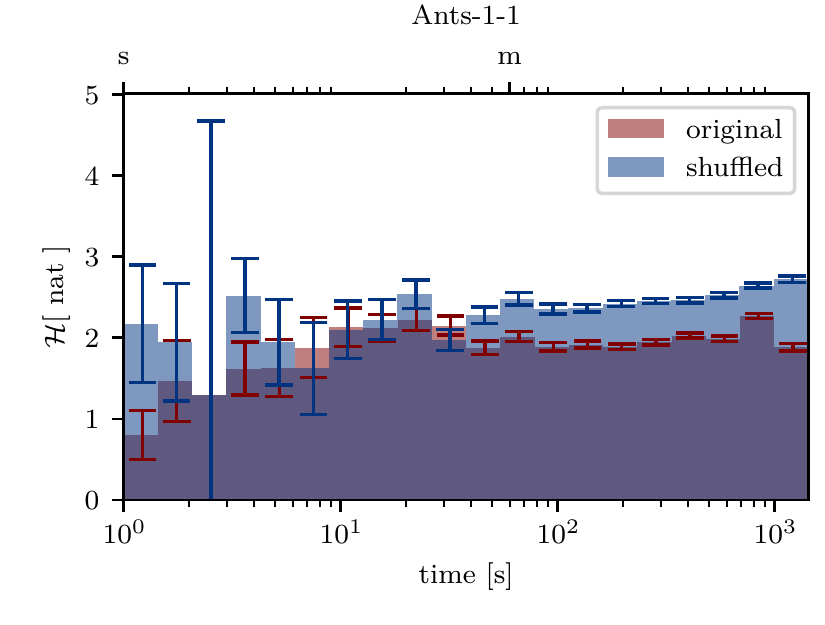}
    \includegraphics[width=0.49\textwidth]{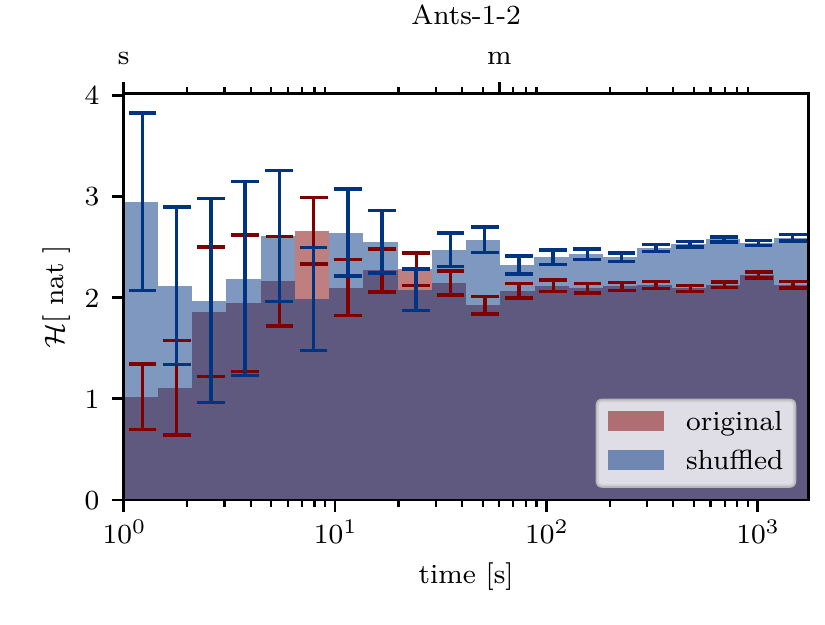}
    \includegraphics[width=0.49\textwidth]{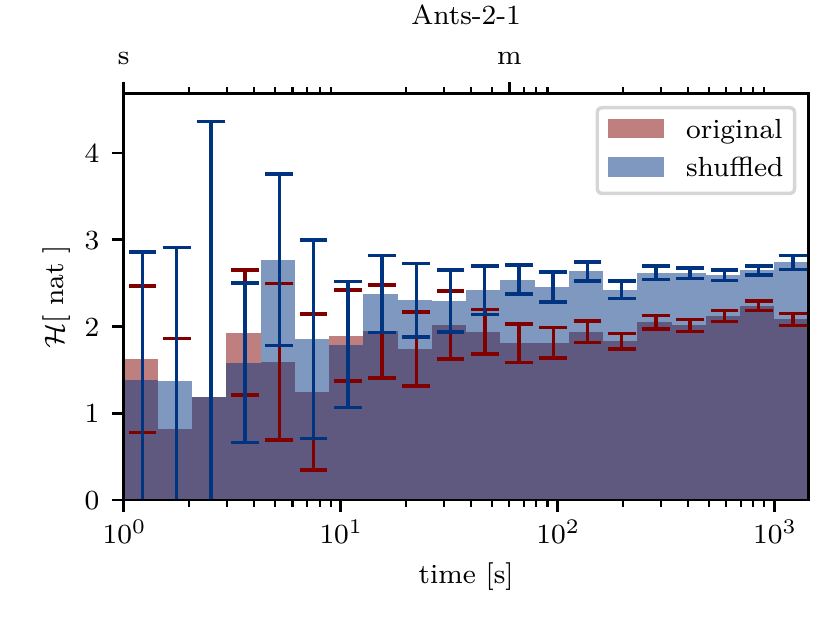}
    \includegraphics[width=0.49\textwidth]{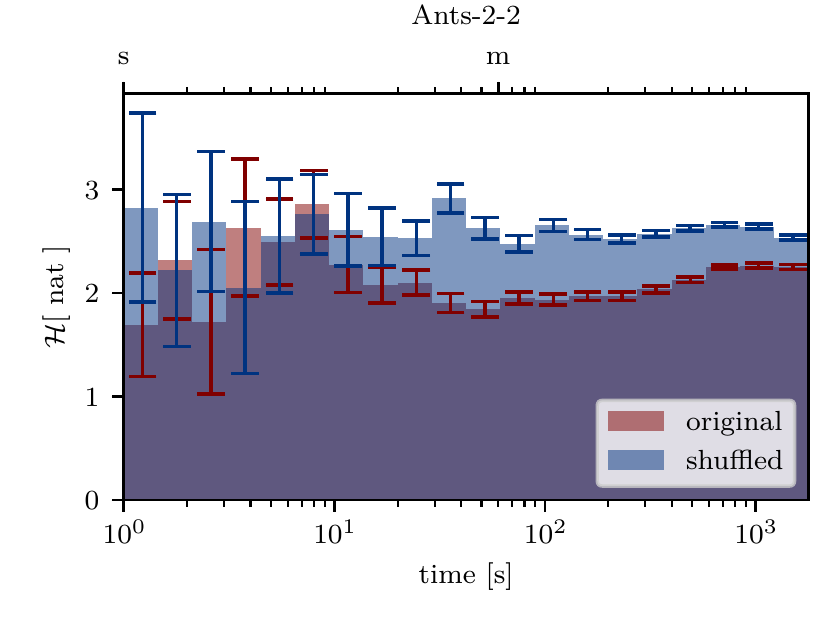}
    \includegraphics[width=0.49\textwidth]{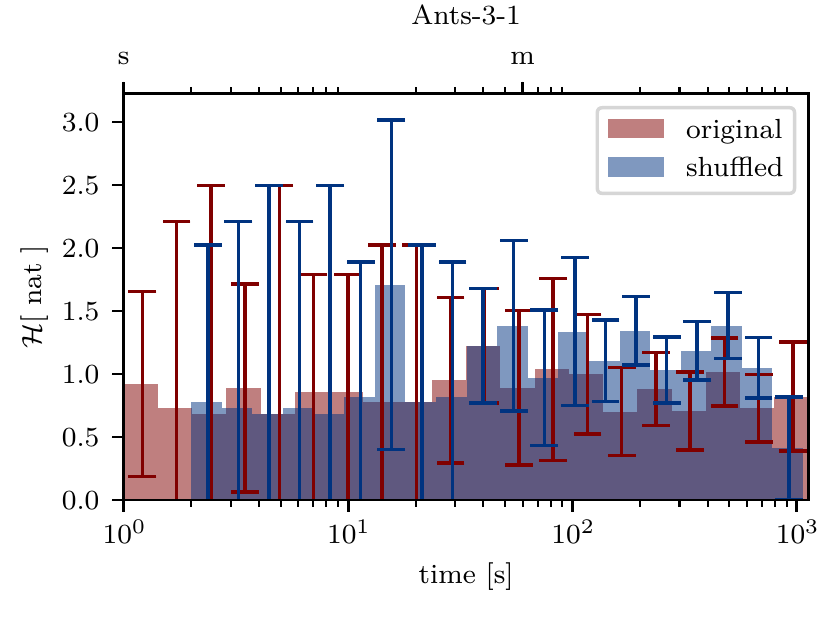}
    \includegraphics[width=0.49\textwidth]{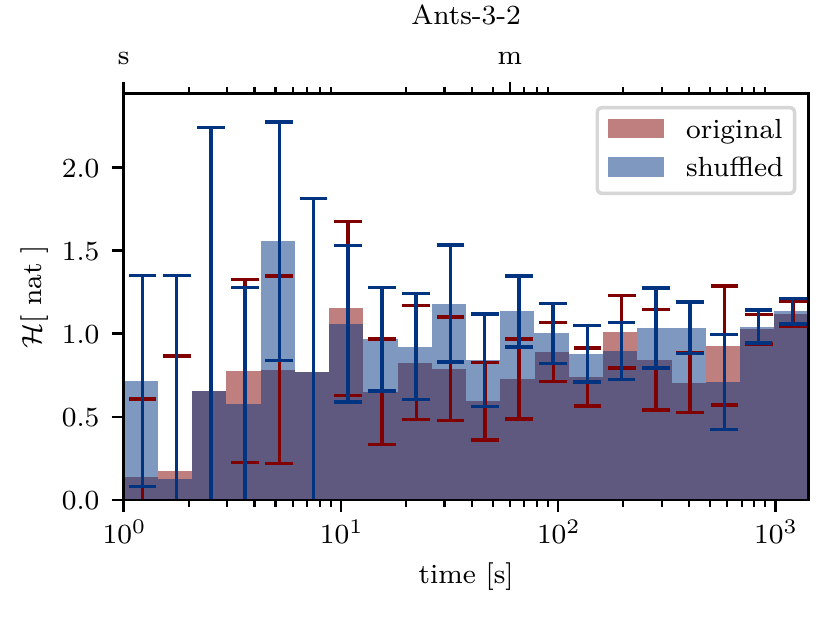}
    \caption{\NameDisorder{} as a function of the timescale $\timewindow$ in temporal networks of antenna contacts in ant collonies. For each temporal network, we show the \nameDisorder{} of the original and of a shuffled network. Timescale $\timewindow$ is represented with the $x$-limits of the bar, and \nameDisorder{} is represented as the height of the bar. Error bars indicate the error of the \nameDisorder{} estimates. }
    \label{tsd:fig:diverse1}
\end{figure*}

\begin{figure*}[!ht]
    \centering
    
    \includegraphics[width=0.49\textwidth]{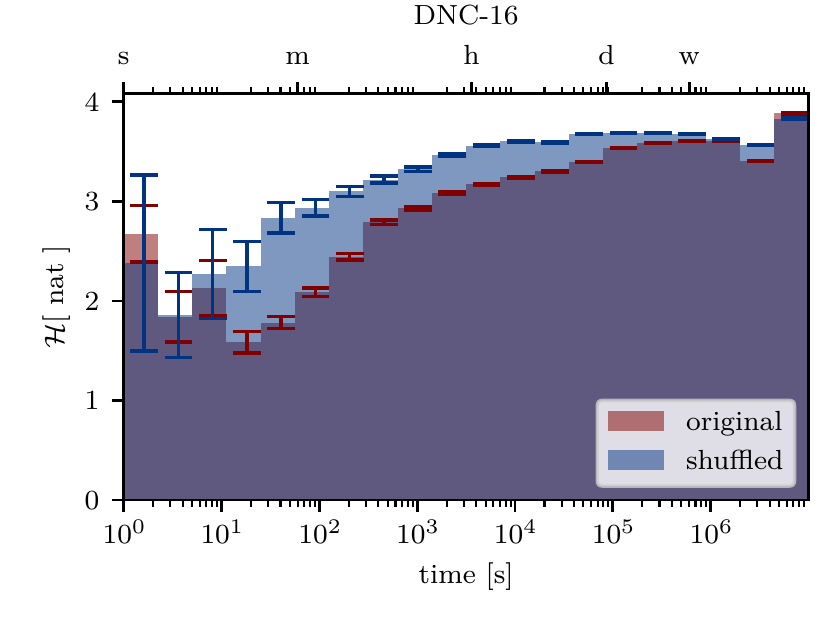}
    \includegraphics[width=0.49\textwidth]{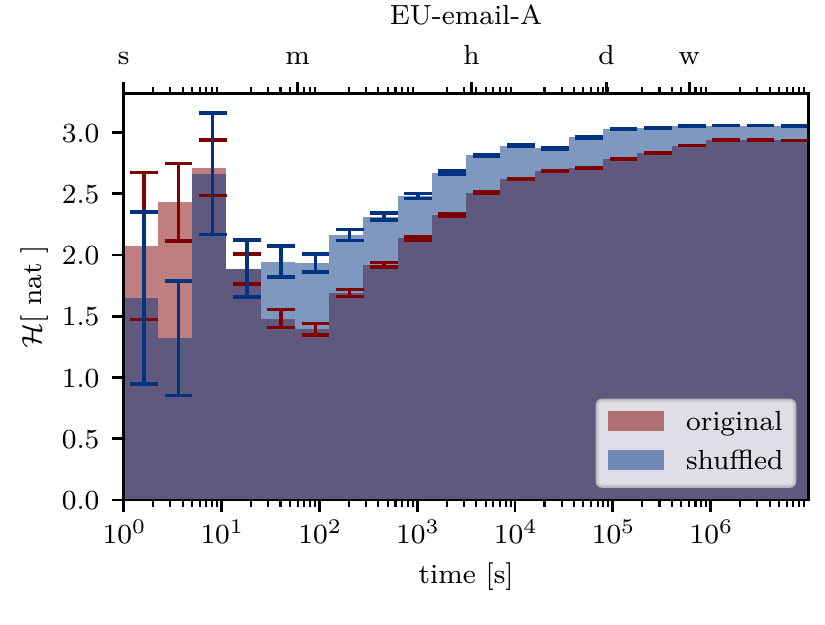}
    \includegraphics[width=0.49\textwidth]{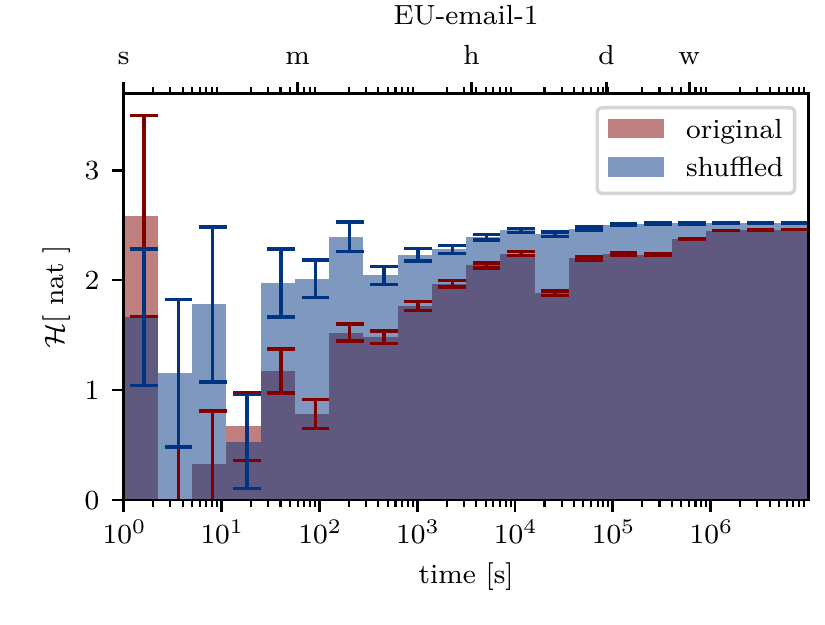}
    \includegraphics[width=0.49\textwidth]{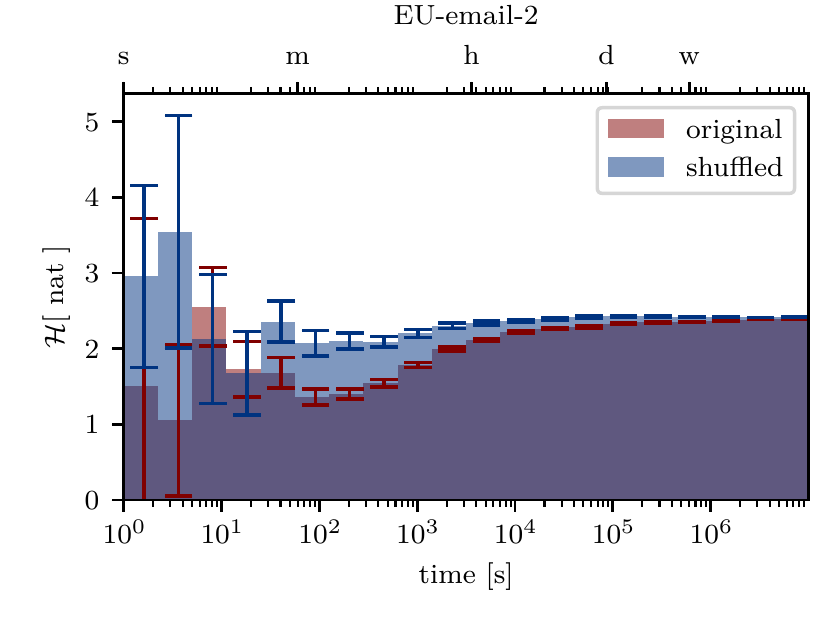}
    \includegraphics[width=0.49\textwidth]{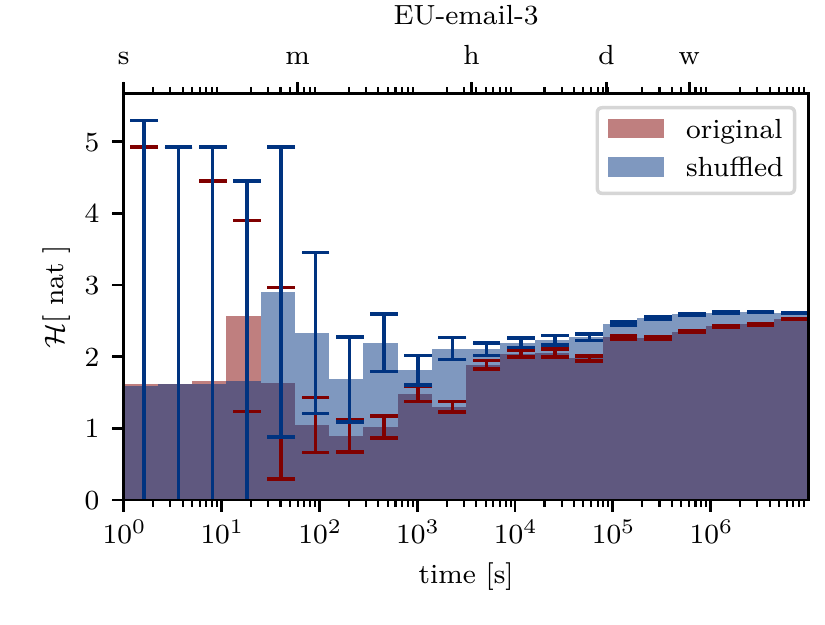}
    \includegraphics[width=0.49\textwidth]{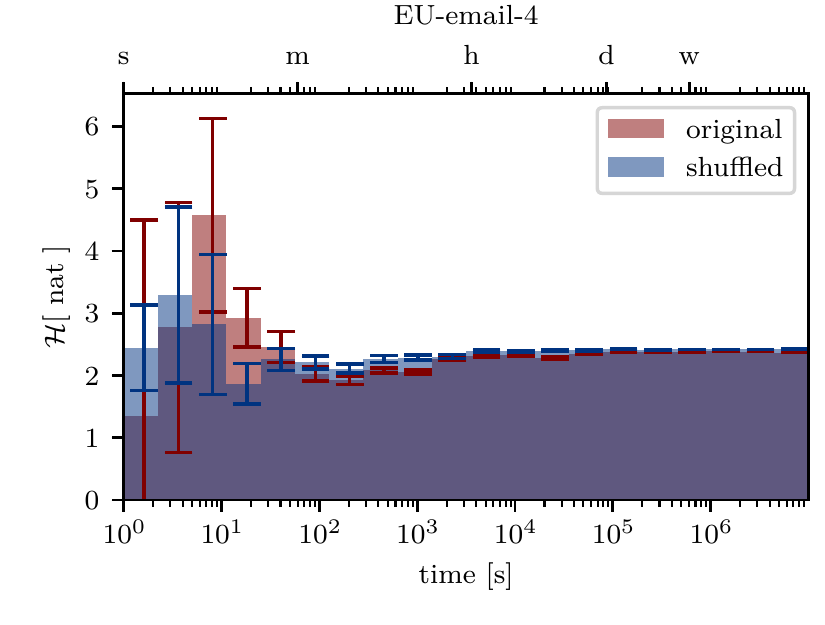}
    \caption{\NameDisorder{} as a function of the timescale $\timewindow$ in temporal networks of email correspondence. For each temporal network, we show the \nameDisorder{} of the original and of a shuffled network. Timescale $\timewindow$ is represented with the $x$-limits of the bar, and \nameDisorder{} is represented as the height of the bar. Error bars indicate the error of the \nameDisorder{} estimates.}
    \label{tsd:fig:diverse2}
\end{figure*}

\begin{figure*}[!ht]
    \centering
    
    \includegraphics[width=0.49\textwidth]{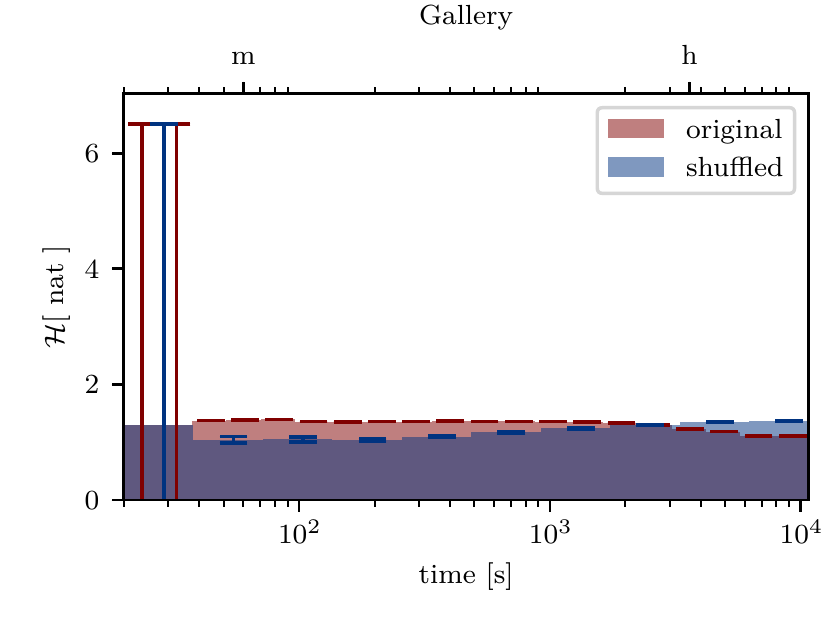}
    \includegraphics[width=0.49\textwidth]{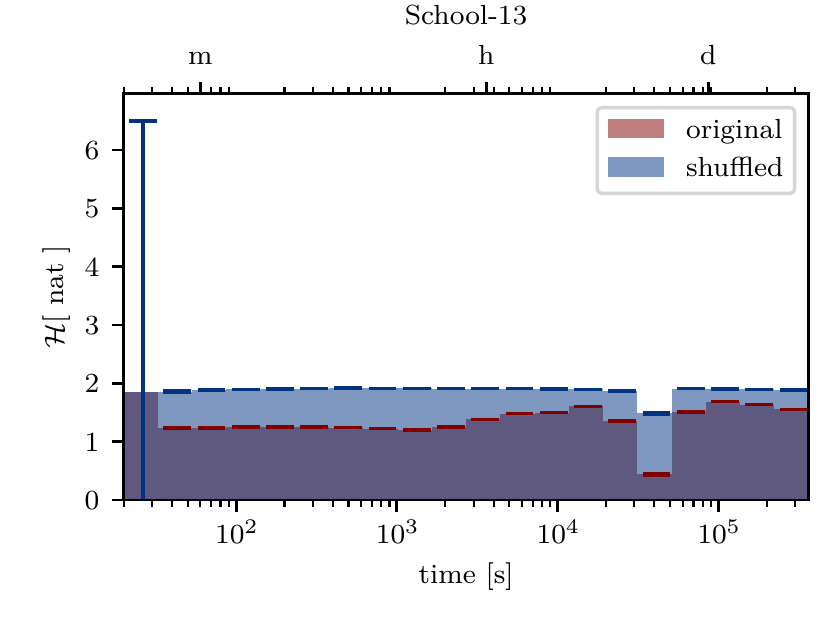}
    \includegraphics[width=0.49\textwidth]{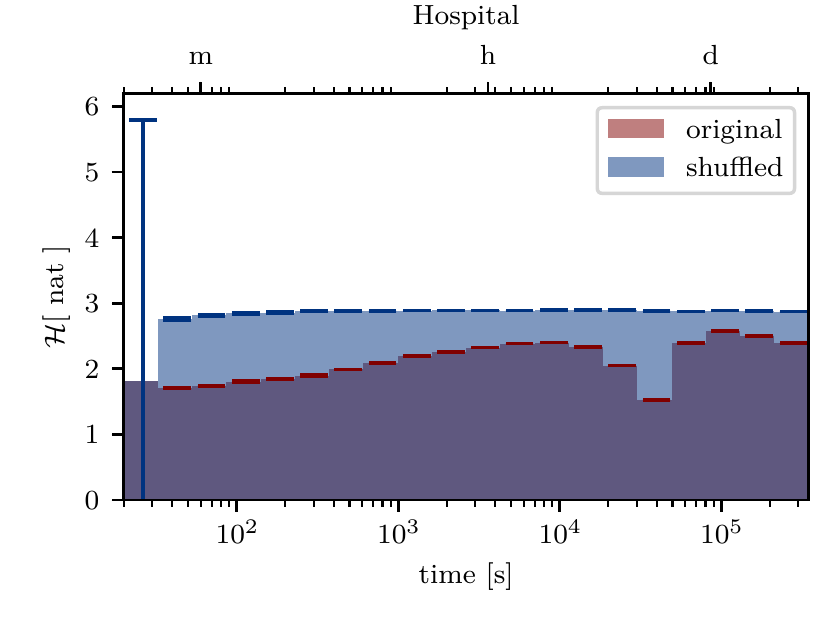}
    \includegraphics[width=0.49\textwidth]{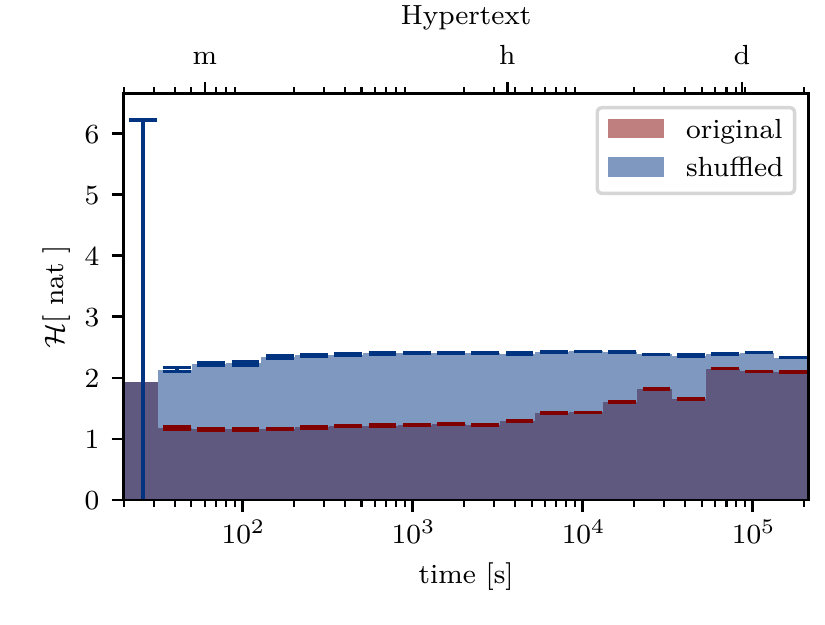}
    \includegraphics[width=0.49\textwidth]{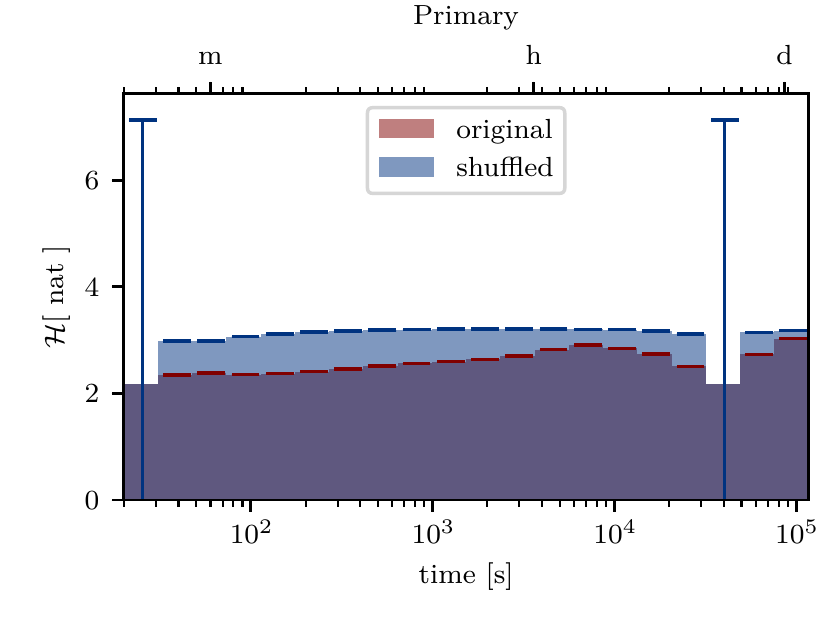}
    \includegraphics[width=0.49\textwidth]{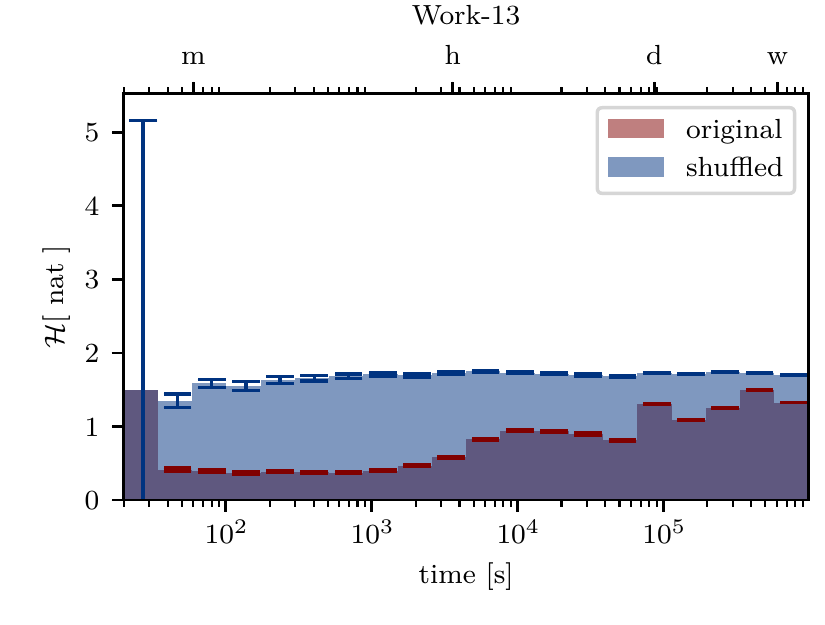}
    \caption{\NameDisorder{} as a function of the timescale $\timewindow$ in temporal networks of human face-to-face interactions measured by the SocioPatterns collaboration. For each temporal network, we show the \nameDisorder{} of the original and of a shuffled network. Timescale $\timewindow$ is represented with the $x$-limits of the bar, and \nameDisorder{} is represented as the height of the bar. Error bars indicate the error of the \nameDisorder{} estimates.}
    \label{tsd:fig:diverse3}
\end{figure*}

\begin{figure*}[!ht]
    \centering
    \includegraphics[]{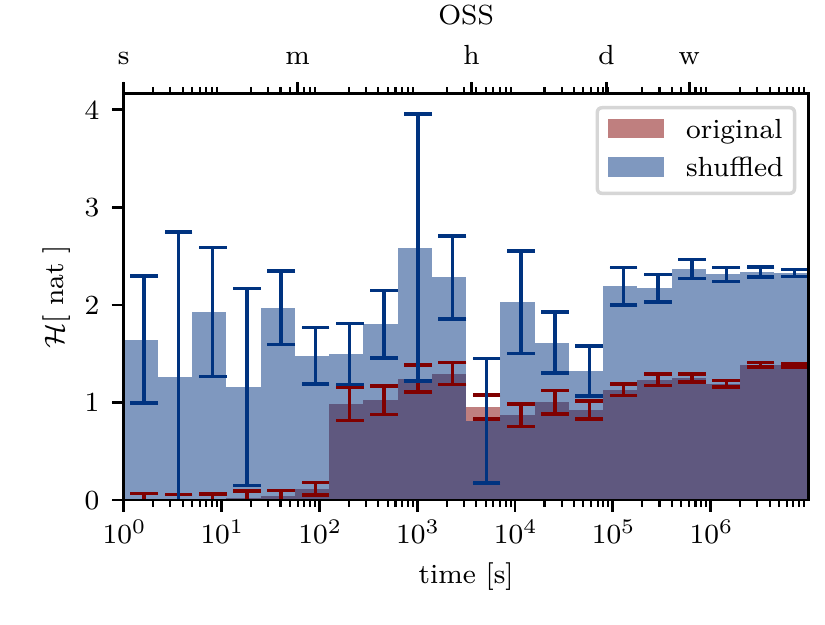}
    \caption{\NameDisorder{} as a function of the timescale $\timewindow$ in temporal networks ASSIGN relationships between members of the Open Source Software community Apache. We show the \nameDisorder{} of the original and of a shuffled network. Timescale $\timewindow$ is represented with the $x$-limits of the bar, and \nameDisorder{} is represented as the height of the bar. Error bars indicate the error of the \nameDisorder{} estimates.}
    \label{tsd:fig:diverse4}
\end{figure*}

\clearpage
\section{Conditional entropy: The chain rule}
For discrete random variables $X$ and $Y$, the definition of the entropy (in nats) is 
\begin{equation}
    H(X) = - \sum_{x} p(X=x) \ln p(X=x) \nonumber
\end{equation}
and the definition of conditional entropy (in nats) $H(Y|X)$ is:
\begin{equation}
    H(Y|X) = - \sum_{x,y} p(X=x,Y=y) \ln\frac{p(X=x,Y=y)}{p(X=x)} \nonumber
\end{equation}
In the following, we use the above definitions to derive the chain rule of conditional entropy:
\begin{align}
    H(Y|X) 
    &= - \sum_{x,y} p(X=x,Y=y)  \left(\ln p(X=x,Y=y) - \ln p(X=x) \right) = \nonumber\\
    &= - \sum_{x,y} p(X=x,Y=y) \ln p(X=x,Y=y) -\left[-\sum_{x,y}p(X=x,Y=y)\ln(p(X=x)))\right] = \nonumber\\
    &= H(X,Y) - \left[-\sum_{x,y} p(Y=y|X=x) p(X=x)\ln(p(X=x)))\right] = \nonumber\\
    &= H(X,Y) - \left[-\sum_{x} p(X=x) \ln(p(X=x))) \cancelto{1}{ \left(\sum_{y}p(Y=y|X=x) \right)} \right] = \nonumber\\
    &= H(X,Y) - H(X).
\end{align}

\end{document}